\pdfoutput=1
\documentclass[11pt]{article}
\usepackage{jcappub}

\usepackage[utf8]{inputenc}

\usepackage[normalem]{ulem}

\usepackage{hyperref}
\usepackage{comment}
\usepackage{amsmath}
\usepackage{amssymb}
\usepackage{xcolor}
\usepackage{physics}
\usepackage{graphicx}

\usepackage[colorinlistoftodos,draft]{todonotes}

\usepackage{pgf}

\numberwithin{equation}{section}
\DeclareMathOperator{\sgn}{sgn}

\begin{document}

\title{A Critique of Covariant Emergent Gravity}
\author{Kirill Zatrimaylov}
\emailAdd{kirill.zatrimaylov@sns.it}
\affiliation{\emph{Scuola Normale Superiore and I.N.F.N.}\\\emph{Piazza dei Cavalieri 7, 56126, Pisa, Italy}}

\abstract{I address some problems encountered in the formulation of relativistic models encompassing the MOND phenomenology of radial acceleration. I explore scalar and vector theories with fractional kinetic terms and $f(R)$-type gravity, demanding that the energy density be bounded from below and that superluminal modes be absent, but also that some consistency constraints with observational results hold. I identify configurations whose energy is unbounded from below and formulate some no-go statements for vector field models and modified gravity theories. Finally, I discuss superfluid dark matter as a hybrid theory lying between CDM and modified gravity, highlighting some difficulties present also in this case, which appears preferable to the others.}

\maketitle
\baselineskip=20pt
\section{Introduction}\label{sec:introduction}

MOND (Modified Newtonian dynamics), proposed by M. Milgrom in 1981~\cite{Milgrom:1983ca}, was meant originally as an alternative to the cold dark matter paradigm: it aims to explain phenomena usually attributed to dark matter via Newtonian laws (either the law of gravity or the law of inertia) that are modified at large distances and small accelerations. Until recently, its actual significance has been somewhat obscure, since it does account for one class of observations (galaxy rotation curves) without addressing other key issues (CMB, primordial structure formation, and displacement between luminous and dark matter components in galaxy cluster collisions). The Bullet Cluster is a notorious example in this respect~\cite{Clowe:2006eq}~\cite{Markevitch:2003at}, although its high collision velocities are also challenging to explain within $\Lambda$CDM~\cite{Lee:2010hja}~\cite{Thompson:2014zra}~\cite{Bouillot:2014hda}, and MOND was actually conjectured to be a better framework for reproducing them~\cite{Angus:2007qj}.

The last years have witnessed a renewed interest in MOND, in two contexts. The first, superfluid dark matter (SfDM), was proposed in~\cite{Berezhiani:2015bqa}. It is a two-phase dark matter model: on cosmological scales it behaves like a light scalar field mimicking cold dark matter, while on galactic scales it condenses in gravitational wells becoming a superfluid. The phonons of the superfluid then produce a dragging force that mimics the effects of MOND.
The second, emergent gravity, is a variant of the ``dark fluid`` approach in which dark matter and dark energy are different manifestations of the same phenomenon~\cite{Verlinde:2016toy}. Dark energy is ascribed to an elastic medium, while the ``MOND force`` is regarded as the medium's response to baryonic matter. Under the spell of the AdS/CFT correspondence, the ``dark fluid`` appears an artifact of quantum gravity, and within a corpuscular approach to quantum gravity it was also interpreted as a Bose-Einstein condensate of gravitons~\cite{Tuveri:2019zor}~\cite{Giusti:2019wdx}.  The initial model of~\cite{Verlinde:2016toy} was only applicable to non-relativistic and spherically symmetric systems, but a relativistic extension, known as covariant emergent gravity (CEG), was later proposed in~\cite{Hossenfelder:2017eoh}. Aside from reproducing MOND, CEG also yields corrections to it and admits a de Sitter vacuum solution that is consistent with the ``dark fluid`` interpretation in~\cite{Verlinde:2016toy}.
 
In this paper I study possible realizations of the MOND regime, taking into account that the non-relativistic ``deep MOND`` equations require a fractional kinetic term in the Lagrangian. This could either remain manifest in the covariant formulation, as in CEG, or could emerge in specific vacua, as in SfDM, or could be completely absent from the relativistic theory, as in $f(R)$ gravity. I subject these three options to a detailed scrutiny, demanding that three key conditions be satisfied:
\begin{itemize}
    \item the field energy density should be bounded from below;
    \item superluminal propagation should be absent in relativistic settings;
    \item the models should be able to account for gravitational lensing.
\end{itemize}
Combining this analysis with previously known results, I attempt to formulate a no-go statement on relativistic extensions of MOND, and comment on its implications for general long-range modifications of gravity.

The paper is structured as follows. In Section~\ref{sec:MOND_introduction} I briefly recall the observational evidence for MOND, its basic postulates and the formulation of the ``deep MOND`` regime associated to a scale invariant action with a fractional kinetic term. In Section~\ref{sec:Relativistic_MG} I address the relativistic generalizations of this action proposed in the literature and show that, with scalar or vector fields, there are generically configurations whose energy density is unbounded from below. In particular, I show that the CEG Hamiltonian is unbounded from below, that vector field theories in general cannot reproduce MOND in a natural fashion, and derive generic conditions for scalar and vector field theories with fractional kinetic terms to have lower bounds for their energy densities. I also consider $f(R)$ gravity and demonstrate that, while it can allow MOND-type potentials, it is problematic to connect their strengths, as one would need, to Newtonian potentials at shorter distances. In Section~\ref{sec:Khoury} I address the more intricate option that the non--relativistic ``deep MOND" Lagrangian describes a broken phase of a more conventional scalar--field Lagrangian. This last setting appears less problematic than the others, and yet I show that it entails a problem with the vacuum energy in the broken phase. I conclude in Section~\ref{sec:conclusions} with a summary of the work, some comments on its potential implications and a discussion of possible future developments.

\section{A Brief Overview of MOND}\label{sec:MOND_introduction}
\begin{figure}
	\label{Tully-Fisher}
	\begin{center}
		\includegraphics[width=50mm]{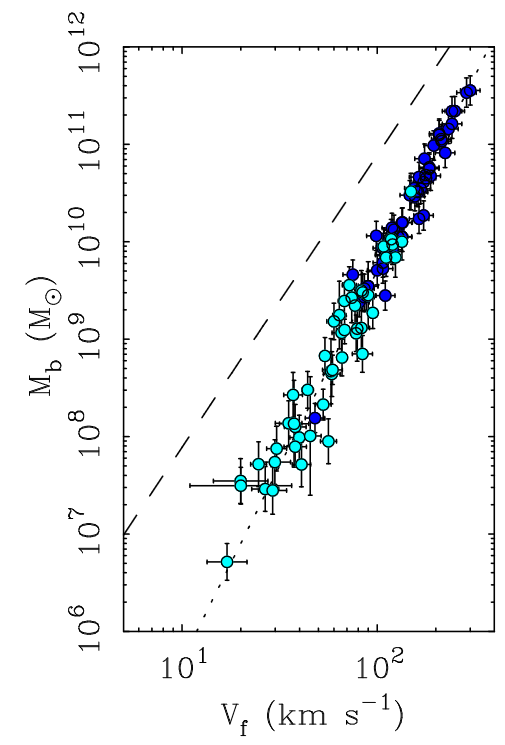}
	\end{center}
	\caption{The baryonic Tully-Fisher relation (BTFR), reproduced from~\cite{Famaey:2011kh}. Circles are data (in particular, dark-blue circles concern star-dominated galaxies, while light-blue circles concern gas-dominated galaxies), the dotted line is a $v^4$ fit, while the dashed line is a $v^3$ fit, as in cosmological collapse theory.}
\end{figure}

While cold dark matter can account for the dynamics at very large scales, at galaxy and galaxy--cluster scales one is confronted with numerous puzzles or discrepancies that could reflect yet unknown physics. One of these puzzles is the so-called Tully-Fisher relation (see fig.~\ref{Tully-Fisher}), according to which the total baryonic mass of galaxies is proportional to the fourth power of the asymptotic velocity (the rotation velocity at the largest observed distances), $M_b\propto v_a^4$.
Since $v_a$ should be almost entirely determined by dark matter, this relation indicates an unexpectedly tight correspondence between the dark and luminous components. One can recover the Tully-Fisher relation considering only baryonic matter and postulating that the actual physical acceleration is related to the Newtonian value by
\begin{equation}
a=a_N\, \nu\left(\frac{a_N}{a_0}\right) \ ,
\end{equation}
where $\nu(x)$ is a function with the asymptotics
\begin{equation}
\nu(x) \sim 1 \quad (x\gg 1) \ ,\qquad
\nu(x) \sim \frac{1}{\sqrt{x}} \quad (x\ll 1) \ ,
\end{equation}
and $a_0$ is a fixed acceleration scale. For accelerations much larger than $a_0$ one thus recovers the standard Newtonian law, while for $a\ll a_0$ (the ``deep MOND" regime) one obtains the so-called radial acceleration relation (RAR)
\begin{equation}
a=\sqrt{a_0a_N} \ ,
\end{equation}
which matches the observational data at large distances~\cite{McGaugh:2016leg} and leads indeed to the Tully-Fisher relation, since
\begin{equation}\label{TF}
\frac{v^2}{r} \ = \ \frac{\sqrt{a_0GM_b}}{r} \ \Longleftrightarrow \ v^4 \ = \ a_0GM_b \ .
\end{equation}
Fits of observational data reveal that $a_0$ is of order $\sqrt{\Lambda}\sim H_0$, a puzzling coincidence since it seems difficult to justify correlations between galaxy scales and the Hubble scale. It should be noted, however, that galaxy rotation curves are not \textit{exactly} flat at large distances. Many of them have a slightly decreasing or even a slightly increasing slope~\cite{Persic:1995ru}~\cite{Salucci:2007tm}, which indicates that the MOND regime is possibly a leading-order approximation of some underlying theory. In addition, RAR was conjectured in~\cite{DiPaolo:2018mae} to be the limiting case of a more general empirical relation, known as the GGBX relation.

One can derive a Lagrangian formulation of MOND noting that the ``deep MOND" dynamics ought to be scale invariant, since eq.~\eqref{TF} does not change under the transformations
\begin{equation}
\vec{r} \ \rightarrow \ \lambda\vec{r} \ , \ t \ \rightarrow \ \lambda t \ . \label{redefinitions}
\end{equation}
Alternatively, the MOND dynamics depends only on the quantity $GMa_0$, whose dimension is $[L/T]^4$, and not on $G$, $M$, and $a_0$ separately~\cite{Milgrom:2008cs}.
A typical action principle for Newtonian mechanics is
\begin{equation}
\mathcal{S}_N \ = \ -\frac{1}{8\pi G}\int dt\,d^3x \ (\vec{\nabla}\Phi)^2-\int dt\,d^3x \ \rho\Phi+\int dt\,d^3x \ \frac{\rho v^2}{2} \ ,
\end{equation}
and in order to arrive at scale-invariant equations, all terms should have identical scaling dimensions. Under eq.~\eqref{redefinitions} the third term scales as $\lambda$ ($\rho\propto\lambda^{-3}$), and $\Phi$ should be scale invariant in order to grant the same scaling dimension to the second term. However, this choice would make the first term scale as $\lambda^2$, and therefore, in order to recover MOND, one should change either the third term or the first one. The first approach, known as modified inertia (MI), still lacks a complete theoretical description, because the modified acceleration is a functional of the whole particle trajectory. This means that the corresponding equations of motion would be \emph{non local} in time, and the Lagrangian would not contain a finite number of derivatives of $\vec{r}$. For this reason, MI will not be considered here, but we invite the reader to consult~\cite{Milgrom:1992hr}~\cite{Milgrom:1998sy}~\cite{Milgrom:2005mc} for more details.
However, within the second approach, known as modified gravity, one has the option of altering the first term in order to make it scale like $\lambda$, according to
\begin{equation}\label{MG}
\mathcal{S}_{MG} \ = \ -\,\frac{1}{12\pi Ga_0}\int dt\,d^3x \ \left((\vec{\nabla}\Psi)^2\right)^{3/2}-\int dt\,d^3x \ \rho\Psi+\int dt\,d^3x \ \frac{\rho v^2}{2} \ .
\end{equation}

This model is known as AQUAL (AQUAdratic Lagrangian), and gives the modified Poisson equation
\begin{equation}\label{MP}
\vec{\nabla}(|\vec{\nabla}\Psi|\vec{\nabla}{\Psi}) \ = \ 4\pi Ga_0\rho \ ,
\end{equation}
which, for a point-like mass
\begin{equation}
\rho(r) \ = \ M\delta(\vec{r}) \ ,
\end{equation}
yields indeed the spherically symmetric solution
\begin{equation}\label{MOND_Pot}
\Psi(r) \ = \ -\sqrt{a_0GM}\ln(r)
\end{equation}
and the radial acceleration~\eqref{TF}. Now, if $\Psi$ is an additional field distinct from $\Phi$, including both of them in the Lagrangian can give the total acceleration
\begin{equation}
a_{tot}(a_N)=a_N+\sqrt{a_Na_0} \ .
\end{equation}
This type of function provides a good fit of observational data for a number of galaxy rotation curves~\cite{Hossenfelder:2018vfs}. Since we already know that a field like $\Phi$ can emerge in General Relativity as a perturbation of the metric tensor $g_{\mu\nu}$, it is natural to try and understand the relativistic origin of $\Psi$. In the following we shall work in a ``mostly-plus'' signature.
\section{Covariant Theories with Fractional Kinetic Terms}\label{sec:Relativistic_MG}
The most straightforward approach to obtain a covariant form of the action \eqref{MG} would be to assume a kinetic term of the form
\begin{equation}
\mathcal{L}_k=\chi^{3/2} \ ,
\end{equation}
where $\chi$ is Lorentz invariant and quadratic in derivatives of $\Psi$. Models with non-standard kinetic terms of the form $F(\chi)$ are generally known as k-essence theories. Although fractional powers in the kinetic term may appear strange from a field theory perspective, their emergence is not uncommon, the simplest known example being the action of a relativistic particle,
\begin{equation}
\mathcal{S}=-m \int \ \ dt \ \ \sqrt{1-\vec{v}^2(t)} \ ,
\end{equation}
which becomes singular when $v$ approaches the speed of light.
Other examples include the Dirac-Born-Infeld action~\cite{Born:1934gh}, which plays a role in String Theory~\cite{Fradkin:1985qd}~\cite{Tseytlin:1999dj}, close to the critical field values, the phononic action of the unitary Fermi gas (UFG)~\cite{Son:2005rv}, and the Hamiltonian of fractional quantum mechanics~\cite{Laskin:2002zz}, although the last two are non-relativistic examples. Insofar as these kinetic terms contain only contributions of the form $(\partial\psi)^2$, they do not spoil the Cauchy problem, because the equations of motion do not include derivatives of order higher than two. Schematically, the EOMs of a Lagrangian proportional to $\chi^\lambda$ read
\begin{equation}
\frac{\delta}{\delta\psi}(\partial\psi)^{2\lambda} \ = \ -2\lambda(2\lambda-1)(\partial\psi)^{2\lambda-2}\partial^2\psi \ .
\end{equation}
It should be noted that a Lagrangian of the form $\chi^\lambda$, with non-integer $\lambda$, is generally defined only for $\chi>0$. The region of negative $\chi$ is not accessible, as is the case for the super--luminal regime of a relativistic particle, and its boundary signals the breakdown of the effective field theory. However, it is also possible, in principle, to try and circumvent this limitation considering kinetic terms of the form $\chi^\alpha|\chi|^\beta$, where $\alpha+\beta=\lambda>0$, $\alpha$ is an integer and $\beta$ is positive. The last requirement does not lead to any loss of generality, since $|\chi|^2=\chi^2$. 

A minimal requirement, which we shall enforce in all cases, posits that the energy density of the field, defined as
\begin{equation}
T_{00}=-\frac{2}{\sqrt{-g}}\frac{\delta}{\delta g^{\mu\nu}}\left(\sqrt{-g}\mathcal{L}_k\right)\Big|_{00} \ ,
\end{equation}
be bounded from below,
and for $\mathcal{L}\propto\chi^\alpha|\chi|^\beta$, with $\alpha+\beta>0$, one would get
\begin{equation}\label{A}
-\left(2\lambda\frac{\delta\chi}{\delta g^{\mu\nu}}\Big|_{00}+\chi\right)\chi^{\alpha-1}|\chi|^\beta \ .
\end{equation}
To derive this expression, we used the identity $|x|=x\,\sgn(x)$ and the fact that terms proportional to $\sgn'(\chi)=\delta(\chi)$ vanish for positive values of $\lambda$, while we are excluding negative values of $\lambda$, which would result in a theory that is unbounded at $\chi=0$. Proceeding along these lines, one can conclude that the choice $\mathcal{L}\propto|\chi|^{\lambda-1}\chi$
can encompass both positive and negative values of $\chi$. In particular, as we shall see shortly, the case of interest for us will be $\lambda=\frac{3}{2}$, and the corresponding Lagrangian allowing both signs for $\chi$ would be 
\begin{equation}
\mathcal{L}\propto \chi\, \sqrt{|\chi|} \ .
\end{equation}

A second, equally important requirement, is causality. Namely, if one perturbs the equations of motion around a stable field configuration, letting
\begin{equation}
\Psi=\Psi_0(\vec{r})+\delta\Psi(t,\vec{r}) \ ,
\end{equation}
the eikonal approximation
\begin{equation}
\delta\Psi=Ae^{i\phi} \ ,    
\end{equation}
where $A$ is a slowly varying amplitude and $\phi$ is a quickly oscillating phase, would yield for the wave vector the dispersion relation
\begin{equation}
\omega=\partial_0\phi \ , \ k_i=\partial_i\phi \ ,
\end{equation}
and the group velocity, defined as
\begin{equation}
v_g \ = \ \frac{\partial\omega}{\partial|\vec{k}|} \ ,
\end{equation}
should not be larger than 1.

Turning to the interaction terms, the baryonic density $\rho$ is the 00-component of the matter stress-energy tensor, so that a relativistic interaction term should be of the form 
\begin{equation}\label{Int}
\mathcal{L}_{int}=h^{\mu\nu}T_{\mu\nu} \ ,
\end{equation}
and the previously defined $\Psi$ should determine the ``effective coupling'' $h_{\mu\nu}$, whose exact form depends on whether $\Psi$ is a scalar, vector, or tensor field. Since $T_{\mu\nu}$ is, up to a coefficient, the functional derivative of the matter action with respect to $g_{\mu\nu}$, introducing an interaction term of the form \eqref{Int} is equivalent, to first order in $h_{\mu\nu}$, to coupling the matter Lagrangian to an effective metric $\tilde{g}_{\mu\nu}$:
\begin{equation}
\mathcal{S}_m=\int \ d^4x \ \sqrt{-\tilde{g}} \ \mathcal{L}_m(\tilde{g}_{\mu\nu}) \ ,
\end{equation}
with
\begin{equation}
\tilde{g}_{\mu\nu}=g_{\mu\nu}-2h_{\mu\nu} \ .
\end{equation}
We shall also demand that the interaction term reproduce the effects of gravitational lensing, and we can now analyze in detail various options.
\subsection{Scalar} 
A scalar field is the simplest option that one can consider. For a theory with a Lagrangian of the form
\begin{equation}
\mathcal{L}=-(\partial^\mu\Psi\,\partial_\mu\Psi)^\lambda \ , \label{scalar_kinetic}
\end{equation}
the energy density is
\begin{equation}
T_{00}=(\partial^\mu\Psi\partial_\mu\Psi)^{\lambda-1}\left[(2\lambda-1)(\dot{\Psi})^2+(\vec{\nabla}\Psi)^2\right] \ .
\end{equation}
The expression within square brackets is positive definite for all $\lambda\ge\frac{1}{2}$, but the prefactor can become negative for even $\lambda$ when
\begin{equation}
\dot{\Psi}^2 \ > \ |\vec{\nabla}\Psi|^2 \ .
\end{equation}
We should also exclude this region when considering non-integer $\lambda$, which means $T_{00}$ is positive definite for all positive and non-even (\emph{i.e.} either odd or non-integer) $\lambda$, including the special value $\frac{3}{2}$ that plays a role in the models of interest. Notice, however, that the overall sign choice for the invariant in eq.~\eqref{scalar_kinetic} is at odds with what happens for the relativistic particle, but is needed to account for the MOND regime.

Alternatively, as we have anticipated, one can do something even more at odds with the particle case, allowing both positive and negative values for
\begin{equation}
\chi \ = \ \partial^\mu\Psi\partial_\mu\Psi \ ,
\end{equation}
which correspond to arbitrary long-wavelength modes, considering the Lagrangian
\begin{equation}
\mathcal{L}=-|\partial^\mu\Psi\partial_\mu\Psi|^{\lambda-1}(\partial^\mu\Psi\partial_\mu\Psi)
\end{equation}
with the energy density
\begin{equation}
T_{00}=|\partial^\mu\Psi\partial_\mu\Psi|^{\lambda-1}\left[(2\lambda-1)(\dot{\Psi})^2+(\vec{\nabla}\Psi)^2\right] \ .
\end{equation}
In this theory the energy density is bounded from below for all $\lambda\ge\frac{1}{2}$ (even values of $\lambda$ are not a problem, since the prefactor cannot be negative, but $\lambda$ cannot be smaller than $\frac{1}{2}$, if we now allow positive and negative values of $\chi$.

However, in all cases to reproduce \eqref{MG} $\Psi$ would couple naturally to the trace of $T_{\mu\nu}$, for instance considering
\begin{equation}\label{D}
\mathcal{L}= -\frac{1}{12\pi Ga_0}(\partial^\mu\Psi\partial_\mu\Psi)^{3/2}-\Psi T^\alpha_\alpha \ .
\end{equation}
As a result, this ``scalar MOND" setting is unable to account for gravitational lensing, since the Maxwell stress-energy tensor is traceless. Moreover, it was proved in~\cite{Bekenstein:2004ne} that ``scalar MOND", alternatively known as RAQUAL (relativistic AQUAL), has superluminal modes.

A more complicated version of scalar MOND, known as TEVES (tensor-vector-scalar gravity), was also proposed in~\cite{Bekenstein:2004ne}. In this setup, $\Psi$ is supplemented by a Maxwell-type vector field $V_{\mu}$, which allows to reproduce gravitational lensing, while also eliminating superluminal propagation. However, because of this vector field, matter and gravity couple to two different (conformally unrelated) metrics, which means the gravitational waves would propagate on geodesics different from those of photons - a possibility that was clearly ruled out by the gravitational--wave observations of LIGO~\cite{Boran:2017rdn}.
\subsection{Vector} A vector field is a more complicated option: a generic vector kinetic term is of the form
\begin{equation}
\mathcal{L}=\chi^\lambda \ ,
\end{equation}
with
\begin{equation}
\chi \ = \ \alpha(\nabla_\alpha u^\alpha)^2+\beta(\nabla_\mu u_\nu)(\nabla^\mu u^\nu)+\gamma(\nabla_\mu u_\nu)(\nabla^\nu u^\mu) \ .
\end{equation}
In components, it is given by
\begin{equation}
\begin{gathered}
\chi \ = \ \alpha(\partial_0u_0-\sum_i\partial_iu_i)^2+\frac{(\beta+\gamma)}{2}\left(2(\partial_0u_0)^2+\sum_{i,j}(\partial_iu_j+\partial_ju_i)^2\right.\\
\left.-\sum_i(\partial_0u_i+\partial_iu_0)^2\right)+\frac{(\beta-\gamma)}{2}\left(\sum_{i,j}(\partial_iu_j-\partial_ju_i)^2-\sum_i(\partial_0u_i-\partial_iu_0)^2\right) \ ,
\end{gathered}
\end{equation}
and
\begin{equation}
\begin{gathered}
\frac{\delta\chi}{\delta g^{\mu\nu}} \ = \ -(3\alpha+\beta+\gamma)(\partial_0u_0)^2 +  (4\alpha+\beta+\gamma)(\partial_0u_0)\sum_i(\partial_iu_i)\\
-\alpha\sum_i(\partial_iu_i)^2-2\alpha\sum_{i\neq j}(\partial_iu_i)(\partial_ju_j)+\frac{(\beta-\gamma)}{2}\sum_i(\partial_iu_0-\partial_0u_i)^2\\
+\frac{(\beta+\gamma)}{2}\sum_i\left((\partial_0u_i)^2-(\partial_iu_0)^2\right)+O(\partial^2u) \ .
\end{gathered}
\end{equation}
To compute the variation of $\chi$, we also took into account contributions involving the Christoffel symbols $\Gamma\propto\partial g$, since after integrating by parts they yield terms that survive even in flat spacetime backgrounds. There are also terms proportional to $u(\partial^2u)$, but the problem manifests itself already with linear field configurations of the type 
\begin{equation}
u_\mu \ = \ A_\mu+B_{\mu\nu}x^\nu \ , \label{w}
\end{equation}
for which they vanish.

It should be noted that these special configurations solve the equations of motion, since the EOMs are proportional to \textit{second-order} derivatives of $u$~\footnote{~\cite{Hossenfelder:2017eoh} and~\cite{Dai:2017guq} propose either a mass term or a quartic self-interaction term for $u_\mu$, but their contribution to the EOMs can be made negligible if we consider very small values of $A_\mu$ and a region of spacetime sufficiently close to the origin, so that the $B_{\mu\nu}x^\nu$ are small even for large enough values of $B_{\mu\nu}$.}. For the configurations in eq.~\eqref{w}, one thus obtains
\begin{equation}
\begin{gathered}
\frac{T_{00}}{2\chi^{\lambda-1}} = \left((3\lambda-\frac{1}{2})\alpha+(\lambda-\frac{1}{2})(\beta+\gamma)\right)(\partial_0u_0)^2\\
+ \ ((1-4\lambda)\alpha-\lambda(\beta+\gamma))(\partial_0u_0)(\sum_i\partial_iu_i)\\
+\ \left((\lambda-\frac{1}{2})\alpha-\frac{1}{2}(\beta+\gamma)\right)\sum_i(\partial_iu_i)^2+(2\lambda-1)\alpha\sum_{i\neq j}(\partial_iu_i)(\partial_ju_j)\\
- \ \frac{\beta}{2}\sum_{i\neq j}(\partial_iu_j)^2-\gamma\sum_{i\neq j}(\partial_iu_j)(\partial_ju_i)+\left(\lambda\gamma+\frac{\beta}{2}\right)(\partial_iu_0)^2\\
+ \ \left(\frac{1}{2}-\lambda\right)\beta(\partial_0u_i)^2+\left(\gamma(1-\lambda)+\lambda\beta\right)\sum_i(\partial_0u_i)(\partial_iu_0) \ .
\end{gathered}
\end{equation}
This quadratic form rests on a 16 x 16 matrix, and the positivity condition for the energy is equivalent to the requirement that all its eigenvalues have the same sign~\footnote{In principle, if they were all negative, one could change the overall sign of the Lagrangian.}. This matrix comprises several blocks: for each $\partial_iu_j$ and $\partial_ju_i$ ($i\neq j$) there is a  2 x 2 block
\begin{equation}
-\frac{1}{2}
\begin{pmatrix}
	\beta & \gamma \\
	\gamma & \beta
\end{pmatrix}
\end{equation}
with eigenvalues
\begin{equation}
\Lambda_{3,4} \ = \ -\frac{1}{2}(\beta\pm\gamma) \ .
\end{equation}
One can introduce the new variables
\begin{equation}
\begin{gathered}
\delta \ = \ \frac{1}{2}(\beta+\gamma) \ , \\
\xi \ = \ \frac{1}{2}(\beta-\gamma) \ ,
\end{gathered}
\end{equation}
and the energy positivity condition requires that $\delta$ and $\xi$ have the same sign.
The other $2 \times 2$ block corresponds to the products of $\partial_0u_i$ and $\partial_iu_0$ (for each i), is
\begin{equation}
\begin{pmatrix}
	\lambda\gamma+\frac{\beta}{2} & \frac{\gamma}{2}(1-\lambda)+\lambda\frac{\beta}{2}\\
	\frac{\gamma}{2}(1-\lambda)+\lambda\frac{\beta}{2} & (1-2\lambda)\frac{\beta}{2}
\end{pmatrix} \ ,
\end{equation}
and its two eigenvalues have the same sign if
\begin{equation}
-\lambda^2\delta^2\ -\ 2\left(\lambda\,-\,\frac{1}{2}\right)\delta\,\xi\ \ge \ 0 \ .
\end{equation}
Since the first term in $b$ is negative definite for $\delta\neq0$, the second should be positive to satisfy the condition. But for $\lambda\ge\frac{1}{2}$, it can only be positive if $\delta\xi<0$, which cannot be true, since $\delta$ and $\xi$ should have the same sign, as we have seen. Therefore the only option is $\delta=0$, \emph{i.e.} $\gamma=-\beta$.

Finally, for products of $\partial_0u_0$ and $\partial_iu_i$ we have the 4 x 4 block:
\begin{equation}
\begin{pmatrix}
	(3\lambda-\frac{1}{2})\alpha+2(\lambda-\frac{1}{2})\delta & -(2\lambda-\frac{1}{2})\alpha-\lambda\delta & -(2\lambda-\frac{1}{2})\alpha-\lambda\delta & -(2\lambda-\frac{1}{2})\alpha-\lambda\delta \\
	-(2\lambda-\frac{1}{2})\alpha-\lambda\delta & (\lambda-\frac{1}{2})\alpha-\delta & (\lambda-\frac{1}{2})\alpha & (\lambda-\frac{1}{2})\alpha \\
	-(2\lambda-\frac{1}{2})\alpha-\lambda\delta & (\lambda-\frac{1}{2})\alpha & (\lambda-\frac{1}{2})\alpha-\delta & (\lambda-\frac{1}{2})\alpha \\
	-(2\lambda-\frac{1}{2})\alpha-\lambda\delta & (\lambda-\frac{1}{2})\alpha & (\lambda-\frac{1}{2})\alpha & (\lambda-\frac{1}{2})\alpha-\delta
\end{pmatrix} \ . \label{e}
\end{equation}
One eigenvalue of this matrix,
\begin{equation}
\Lambda \ = \ -\ \delta \ ,
\end{equation}
is doubly degenerate, since inserting it the second, third, and fourth rows coincide. If one now sets $\delta$ to zero, as required by the preceding discussion, the remaining two eigenvalues are determined by the quadratic equation
\begin{equation}
\Lambda^2 \ - \ 2(3\lambda-1)\alpha\,\Lambda\  - \ 3\lambda^2\alpha^2 \ = \ 0 \ .
\end{equation}
Following the same logic as in the previous case, one should require that the last term be non-negative for the two eigenvalues to have the same sign, but this is not possible unless $\alpha=0$. In conclusion, one is left is Maxwell's choice of parameters $\alpha=0, \beta=-\gamma$. All other options, including the choice made in CEG ($\alpha=\frac{4}{3}, \beta=\gamma=-\frac{1}{2}$)~\cite{Hossenfelder:2017eoh}, result in $T_{00}$ unbounded from below within the class of configurations that we have explored.

Since the relevant quantity is not $T_{00}$ but $\frac{T_{00}}{\chi^{\lambda-1}}$, this analysis suffices only for odd integer values of $\lambda$ (including the Maxwell case $\lambda=1$). If, instead, $\lambda$ is either even or non-integer (the latter is relevant for CEG), one should check whether $\frac{T_{00}}{\chi^{\lambda-1}}$ can become negative in the region where $\chi$ is positive. Otherwise, for even $\lambda$, the sign change would be compensated by the sign change of $\chi^{\lambda-1}$, and $T_{00}$ would remain positive, while for non-integer $\lambda$, regions with negative $\chi$ are simply removed from the configuration space. However, it is possible to construct field configurations that satisfy the following conditions:
\begin{equation}\label{v}
\begin{gathered}
\frac{\delta\chi}{\delta g^{\mu\nu}}\Big|_{00}\ge 0 \ , \\
\chi\ge 0 \ .
\end{gathered}
\end{equation}
For example, for negative $\beta$ and $\gamma$ (as is the case in CEG), a static configuration with $u_0$ as the only non-zero component,
\begin{equation}
u_0 \ = \ A+B_{\mu}x^\mu \ ,\\
u_i \ = \ 0
\end{equation}
would lead to a negative $T_{00}$,
\begin{equation}
T_{00} \ \propto \ \left(\beta+2\lambda\gamma\right)(\sum_i(B_i)^2)\chi^{\lambda-1} \ ,
\end{equation}
while
\begin{equation}
\chi \ = \ -\beta\sum_i(B_i)^2
\end{equation}
would be positive. The coefficients $B_i$ can be arbitrarily large, and this means the theory is not bounded from below.

One can take this argument one step further, supplementing it with the condition
\begin{equation}
\beta\,+\,\gamma \ = \ -\,\frac{3}{4}\,\alpha \ ,
\end{equation}
which was derived in~\cite{Hossenfelder:2017eoh} demanding stress-energy conservation in a de Sitter background. Substituting this condition into $\chi$ and $\frac{\delta\chi}{\delta g^{\mu\nu}}\Big|_{00}$, one can see that the conditions \eqref{v} are both satisfied when $u_0$ is only time-dependent for positive $\alpha$, and when it is only space-dependent for negative $\alpha$ ($u_i=0$ in both cases). Our original statement regarding the eigenvalues was only valid for $\lambda\ge\frac{1}{2}$, but this argument is correct regardless of $\lambda$. However, for the Maxwell choice of parameters, one can see that  
\begin{equation}
\begin{gathered}
\chi \ = \ \gamma\left(\sum_i(\partial_0u_i-\partial_iu_0)^2-\sum_{i,j}(\partial_iu_j-\partial_ju_i)^2\right) \ ,\\
T_{00} \ \propto \ \frac{\gamma}{2}\sum_{i,j}(\partial_iu_j-\partial_ju_i)^2+\gamma(\lambda-\frac{1}{2})(\partial_iu_0-\partial_0u_i)^2 \ ,
\end{gathered}
\end{equation}
which means that the positivity condition is only satisfied for $\lambda\ge\frac{1}{2}$. Following a similar logic, one can also see that theories with even $\lambda$ ($\lambda=2n$) are problematic, even for Maxwell's choice of parameters. Since $\frac{T_{00}}{\chi^{\lambda-1}}$ is positive definite, negative values of $\chi$, \emph{i.e.} such that
\begin{equation}
\begin{gathered}
(\partial_iu_j-\partial_ju_i)^2 \ > \ (\partial_0u_i-\partial_iu_0)^2
\end{gathered}
\end{equation}
would result in negative values of $T_{00}$.
Therefore our conclusion is that the only consistent vector field theories rest on Maxwell's choice of parameters ($\alpha=0, \beta=-\gamma$), and on odd or non-integer values of $\lambda$, with $\lambda\ge\frac{1}{2}$.
Moreover, for the Lagrangian based on $|\chi|^{\lambda-1}\chi$, which can also be defined for negative values of $\chi$, $\lambda$ can be even-valued, but again it cannot be smaller than $\frac{1}{2}$.

We can now address the causality issue. The equations of motion for a Lagrangian of the form
\begin{equation}\label{LagrMaxMOND}
\mathcal{L} \sim \ (D_{\mu\nu}D^{\mu\nu})^\lambda \ ,
\end{equation}
with
\begin{equation}
D_{\mu\nu}=\nabla_\mu u_\nu-\nabla_\nu u_\mu \ ,
\end{equation}
are 
\begin{equation}\label{EQS}
\partial_\mu\Big[D^{\mu\nu}(D_{\alpha\beta}D^{\alpha\beta})^{\lambda-1}\Big]\ =\ J^\nu \ ,
\end{equation}
and in absence of sources reduce to
\begin{equation}
(D_{\mu\nu}D^{\mu\nu})^{\lambda-1}\left[\partial^\mu D_{\mu\nu}+2(\lambda-1)\frac{D_{\mu\nu}D^{\gamma\delta}}{(D_{\alpha\beta}D^{\alpha\beta})}\partial^\mu D_{\gamma\delta}\right] \ = \ 0 \ .
\end{equation}
A perturbation $\delta u$ of a static background configuration of the form
\begin{equation}
u_0=\Psi(\vec{r}) \ , \qquad u_i=0
\end{equation}
in the eikonal approximation and in the Lorenz gauge
\begin{equation}
k_\mu \delta u^\mu=0 \ ,
\end{equation}
yields
\begin{equation}
\begin{gathered}
(\vec{k}^2-\omega^2)\delta u_0+2(\lambda-1)(\vec{n}\vec{k})\left((\vec{n}\vec{k})\delta u_0-\omega(\vec{n}\delta\vec{u})\right)=0 \ ,\\
(\vec{k}^2-\omega^2)(\vec{n}\delta \vec{u})+2(\lambda-1)\omega\left((\vec{n}\vec{k})A_0-\omega(\vec{n}\vec{A})\right)=0 \ , \label{vector_caus}
\end{gathered}
\end{equation}
where
\begin{equation}
n_i=\frac{\partial_i\Psi}{\sqrt{(\partial_j\Psi)^2}}
\end{equation}
is a spacelike vector. Combining eqs.~\eqref{vector_caus} gives a quadratic equation for $\omega^2$, with solutions
\begin{equation}
\omega_1^2=\vec{k}^2 \ , \qquad \omega_2^2=\frac{\vec{k}^2+2(\lambda-1)(\vec{n}\vec{k})^2}{2\lambda-1} \ .
\end{equation}
This means that the group velocity can vary between 1 and $\frac{1}{\sqrt{2\lambda-1}}$, and therefore theories with $\lambda\ge1$ have no problems with causality.

In order to try and reproduce the MOND regime with a vector field, one would be tempted to introduce a coupling of the form
\begin{equation}\label{INT}
\mathcal{L}_{int}\ \propto \ \frac{u_\mu u_\nu}{u}\,T^{\mu\nu} \ ,
\end{equation}
as in~\cite{Hossenfelder:2017eoh}. However, this choice breaks the gauge invariance of $u_\mu$, and is thus inconsistent with the equations \eqref{EQS}, once one selects the Maxwell form, which grants a lower bound on the energy density, as we have seen. The current should be conserved ($\partial_\mu J^\mu$), and this condition is generally not satisfied for 
\begin{equation}
J^\mu\ = \ \frac{\delta\mathcal{L}_{int}}{\delta u_{\mu}}\ \propto \ 2\, \frac{u_\nu}{u}T^{\mu\nu}\ +\ \frac{u^\mu u_\alpha u_\beta}{u^3}T^{\alpha\beta} \ .
\end{equation}
Gauge invariance is guaranteed, however, if the coupling term only involves $D_{\mu\nu}$. Let us therefore consider the class of couplings
\begin{equation}
\mathcal{L}_{int}\ = \ F\left(-D_{\gamma\delta}D^{\gamma\delta}\right)D_\mu^\alpha D_{\nu\alpha}T^{\mu\nu} \ , \label{sources1}
\end{equation}
where the function $F$, which we allow out of despair to recover MOND, will be specified shortly.
For a static point mass, one would look for a spherically symmetric field configuration
\begin{equation}\label{Anz}
u_0=\Psi(r) \ , \ u_i=0
\end{equation}
and the emergence of a MOND--like potential would demand, for consistency, that
\begin{equation}\label{LogPot}
F(\Psi'^2)\ \Psi'^2\propto\ln(r) \ ,
\end{equation}
since the source accompanying $T_{\mu\nu}$ in this case ought to play the role of a scalar potential. Away from a point source the Lagrangian of eq.~\eqref{LagrMaxMOND} would yield the field equation
\begin{equation}
\partial_r(r^2(\Psi')^{2\lambda-1})=0 \ \Rightarrow \ \Psi'=Cr^{\frac{2}{1-2\lambda}} \ ,
\end{equation}
and the purported scaling symmetry demands that $\lambda=\frac{3}{2}$.
The sought logarithmic potential \eqref{LogPot} and gauge invariance would conspire into a non--local dressing for the source coupling of the type
\begin{equation}
F(x)\ =\ \frac{\ln(x^2)}{x^2} \ \Rightarrow\  \mathcal{L}_{int}\propto T^{\mu\nu}\frac{D_\mu^\rho D_{\nu\rho}}{D_{\alpha\beta}D^{\alpha\beta}}\ln\left(-D_{\gamma\delta}D^{\gamma\delta}\right) \ ,
\end{equation}
which appears indeed rather baroque, a substantial overkill.
\subsection{Tensor}
If the ``MOND field" were a tensor, the simplest option would be to use the metric $g_{\mu\nu}$~\cite{Bernal:2011qz}. The requirements of energy positivity and of the absence of Ostrogradsky instability demand that the kinetic term involve only the Ricci scalar $R$~\cite{Ostrogradsky:1850fid}~\cite{Woodard:2006nt}~\cite{Stelle:1977ry}, so that this type of realization of ``tensor MOND" would rest on an action of the type
\begin{equation}
\mathcal{S} \ = \ -\frac{1}{16\pi G} \ \int \ d^4x \ \sqrt{-g} \ f(R) \ ,
\end{equation}
and would thus be a special kind of $f(R)$ gravity.

Since $R$ contains terms of the form $\partial^2g$, one cannot recover directly the AQUAL action \eqref{MG}. Nonetheless, we can verify whether the model can produce a logarithmic potential for a point mass source. Outside the source, the equations of motion for f(R) gravity are
\begin{equation}
\begin{gathered}
f'(R)R_{\mu\nu} \ - \ \frac{1}{2}f(R)g_{\mu\nu} \ + \ (g_{\mu\nu}g^{\alpha\beta} \ - \ \delta_\mu^\alpha\delta_\nu^\beta)\left(f'''(R)\partial_\alpha R\partial_\beta R \right. \\
\left. \ + \ f''(R)\partial_\alpha\partial_\beta R-f''(R)\Gamma^\gamma_{\alpha\beta}\partial_\gamma R\right) \ = \ 0 \ .
\end{gathered}
\end{equation}
For a static and spherically symmetric system, they can be written in the form
\begin{equation}\label{f}
\begin{gathered}
f'R_{00}-\frac{1}{2}fg_{00}+g_{00}g^{rr}\left(f'''(R'(r))^2 \ + \ f''R''(r)\right) \\ 
- \ f''R'g_{00}(g^{rr}\Gamma^r_{rr} \ + \ g^{\theta\theta}\Gamma^r_{\theta\theta} \ + \ g^{\phi\phi}\Gamma^r_{\phi\phi}) \ = \ 0 \ , \\
f'R_{rr} \ - \ \frac{1}{2}fg_{rr} \ - \ f''R'g_{rr}(g^{00}\Gamma^r_{00} \ + \ g^{\theta\theta}\Gamma^r_{\theta\theta} \ + \ g^{\phi\phi}\Gamma^r_{\phi\phi}) \ = \ 0 \ , \\
f'\left(g_{\theta\theta}R_{00} \ - \ g_{00}R_{\theta\theta}\right) \ + \ f''R'\left(g_{\theta\theta}\Gamma^r_{00}-g_{00}\Gamma^r_{\theta\theta}\right) \ = \ 0 \ .
\end{gathered}
\end{equation}
The first equation is the $tt$-component, the second is the $rr$-component, and the third is the combination of $tt$- and $\theta\theta$-components. It is also useful to write the trace equation,
\begin{equation}\label{eqtrace}
f'R \ - \ 2f \ + \ 3\Box f' \ = \ 0 \ ,
\end{equation}
although it is not independent of the three above.

Assuming time-independence and spherical symmetry, one can use Schwarzschild coordinates, letting
\begin{equation}
ds^2 \ = \ -(1+2\Phi)dt^2 \ + \ (1+2\Psi)dr^2 \ + \ r^2(d\theta^2 \ + \ \sin^2\theta d\phi^2) \ . 
\end{equation}
To first order in $\Phi$ and $\Psi$, the Ricci tensor and its trace are then
\begin{equation}
\begin{gathered}
R_{00} \ = \ \Phi'' \ + \ \frac{2}{r}\Phi' \ , \ R_{rr} \ = \ -\Phi'' \ + \ \frac{2}{r}\Psi' \ , \\
R_{\theta\theta} \ = \ 2\Psi \ + \ r(\Psi' \ - \ \Phi') \ , \ R_{\phi\phi} \ = \ \sin^2\theta R_{\theta\theta} \ , \\
R \ = \ -2\Phi'' \ + \ \frac{4}{r}(\Psi' \ - \ \Phi') \ + \ \frac{4}{r^2}\Psi \ ,
\end{gathered}
\end{equation}
while the relevant Christoffel symbols are
\begin{equation}
\begin{gathered}
\Gamma^r_{00} \ = \ \Phi' \ , \ \Gamma^r_{rr} \ = \ \Psi' \ , \ \Gamma^r_{\theta\theta} \ = \ -r \ , \ \Gamma^r_{\phi\phi} \ = \ -r\sin^2\theta \ .
\end{gathered}
\end{equation}
For $f=R^\lambda$, the last equation in \eqref{f} demands that, to leading order in $\Phi$ and $\Psi$, $R'=0$. Now, if one wanted that $\Phi$ be the MOND potential, the preceding equations determine a corresponding form for $\Psi$,
\begin{equation}
\Phi(r) \ = \ C \,\ln(r) \ ,
\end{equation}
\begin{equation}
\begin{gathered}
\Psi \ = \ \frac{C}{2} \left( 1 \ + \ \frac{k_1}{r} \ + \ k_2r^2 \right) \ , \qquad  R \ = \ 12\,k_2\,C \ .
\end{gathered}
\label{lambdaless}
\end{equation}
For $\lambda >1$, the last term, proportional to $k_2$, would not be present, and the special case $k_1=k_2=0$ of this result reproduces the one obtained in~\cite{Mendoza:2015una}. Since the MOND solution belongs to the large family defined by the condition $R=0$, it exists for \emph{all} $\lambda>1$. In particular, the choice  $\lambda=\frac{3}{2}$, suggested by an order-of-magnitude approach and by scaling invariance, as for scalar and vector fields, is in principle possible, as in~\cite{Bernal:2011qz}. At any rate,  $C$ is our expansion parameter in perturbation theory, and therefore one should demand that
\begin{equation}
\left|C \right| \ \ll \ 1 \ .
\end{equation}
On the other hand, $C$ enters the MOND term $\sqrt{GMa_0}$, so that the preceding condition translates into the upper bound
\begin{equation}
M \ \ll \ M_P\frac{L_H}{l_P} \ \sim \ 10^{23} M_\odot \ .
\end{equation}
This is well above typical galaxy masses, which are of the order $10^{10}-10^{12} M_\odot$, so that that the MOND approximation appears justified, in this context, on galaxy scales.

However, it is less evident how to ``glue" the MOND solution to the Schwarzschild one. We have in mind two possible ways to do it: the first is to select a function $f(R)$ that converges to $R$ at small radii and to $R^\lambda$ at large radii, and the second is to use the same $f(R)\propto R^\lambda$ at all distances, while resorting to the \emph{solution} $\Phi$ that combines the MOND potential and Schwarzschild potentials. 
The first approach appears problematic, since $R=0$ in both the Schwarzschild and MOND regimes. Hence, even if one chooses $k_2\neq0$ in order to have a nonzero Ricci scalar in the MOND regime, it seems unclear why at larger distances $\Phi$ and $\Psi$ would ``choose" to converge from the Schwarzschild solution with $R=0$ to the MOND solution with $R\neq0$, even leaving aside the need for the particular MOND--value of $C$, $\sqrt{GMa_0}$, and we do not see how to ascertain whether any of the solutions possess attractor properties. 
The second approach is more viable, because the equation $R'=0$ is linear in $\Phi$ and $\Psi$ at first order, so that any combination of  the Schwarzschild potential and MOND potentials,
\begin{equation}\label{a}
\begin{gathered}
\Phi \ = \ C_1 \ln(\frac{r}{r_0}) \ + \ \frac{C_2}{r} \ ,\\
\Psi \ = \ \frac{C_1}{2}(1 \ + \ \frac{k_1}{r} \ + \ k_2r^2) \ - \ \frac{C_2}{r} \ .
\end{gathered}
\end{equation}
is also a solution, where $r_0$ is an arbitrary length scale. However, how can one link the constant to the singular behavior as $r\to 0$? Namely, supplementing eq.~\eqref{eqtrace} with a source $\rho=M\delta(\vec{r})$ and integrating it yields, to leading order,
\begin{equation}\label{v2}
(r^2R^{\lambda-1})\Big|_{r=0} \ = \ \frac{2}{3(\lambda-1)}GM \ .
\end{equation}
But the MOND potential is sub--dominant with respect to the Schwarzschild term, and therefore there is apparently no way to obtain the condition $C_2=GM$.

In addition, it has been demonstrated in~\cite{Soussa:2003sc} that under a number of reasonable assumptions (such as the stability of gravitational theory), $f(R)$ gravity cannot account for gravitational lensing. This can be easily understood from the fact that $f(R)$ theories are generally equivalent to scalar-tensor theories of the form~\cite{Sotiriou:2008rp}
\begin{equation}
\mathcal{S} \ = \ - \int \ d^4x \ \sqrt{-\tilde{g}} \ \left(\frac{1}{2\kappa}\tilde{R}\ + \ \frac{1}{2}\tilde{g}^{\mu\nu}\partial_\mu\phi\partial_\nu\phi \ + \ U(\phi) \right) \ + \ \int \ d^4x \ \sqrt{-g} \ \mathcal{L}(g_{\mu\nu}) \ ,
\end{equation}
where
\begin{equation}
\tilde{g}_{\mu\nu} \ = \ e^{\sqrt{\frac{2}{3}\kappa}\phi}g_{\mu\nu} \ , \ U(\phi) \ = \ \frac{Rf' \ - \ f}{2\kappa f'^2} \ , \ e^{\sqrt{\frac{2}{3}\kappa}\phi} \ = \ f'(R) \ ,
\end{equation}
and $\kappa=8\pi G$. This means that, in terms of matter coupling, $f(R)$ MOND is basically equivalent to ``scalar MOND". In particular, for $f=KR^\lambda$, the potential is
\begin{equation}
U(\phi) \ = \ \frac{\lambda-1}{2\kappa}K^{\frac{1}{1-\lambda}}\lambda^{\frac{\lambda}{1-\lambda}}\exp(\frac{2-\lambda}{\lambda-1}\sqrt{\frac{2}{3}\kappa}\phi) \ .
\end{equation}
On the scalar-tensor side, one can also try to define a MOND-like solution. Outside the source, the equation of motion for $\phi$ is
\begin{equation}
\frac{1}{r^2}\partial_r(r^2\partial_r\phi) \ = \ \frac{2-\lambda}{\sqrt{6\kappa}}K^{\frac{1}{1-\lambda}}\lambda^{\frac{\lambda}{1-\lambda}}\exp(\frac{2-\lambda}{\lambda-1}\sqrt{\frac{2}{3}\kappa}\phi) \ ,
\end{equation}
and therefore, letting
\begin{equation}
\phi \ = \ C\ln(r) \ ,
\end{equation}
leads to
\begin{equation}
\frac{C}{r^2} \ \propto \ r^{\frac{2-\lambda}{\lambda-1}\sqrt{\frac{2}{3}\kappa}C} .
\end{equation}
For consistency, one should require 
\begin{equation}
C \ = \ \frac{\lambda-1}{2-\lambda}\sqrt{\frac{3}{2\kappa}} \ ,
\end{equation}
and in order to reproduce the MOND prefactor $C=\sqrt{GMa_0}$ one should make $\lambda$ a function of the mass M. This feature was also observed in~\cite{Saffari:2007xc}, and while it allows to bypass the no-go theorem on gravitational lensing, it makes the theory somewhat baroque and ill-defined, for general matter Lagrangians.
Alternatively, for $f(R)=R+KR^\lambda$, the potential is
\begin{equation}
U(\phi) \ = \ \frac{\lambda-1}{2\kappa}K^{\frac{1}{1-\lambda}}\lambda^{\frac{\lambda}{1-\lambda}}e^{-2\sqrt{\frac{2}{3}\kappa}\phi}(e^{\sqrt{\frac{2}{3}\kappa}\phi}-1)^{\frac{\lambda}{\lambda-1}} \ ,
\end{equation}
but the structure of the equation is the same as in the previous case, and suffers from the same problem.
We should note, however, that the field $\phi$, usually known as ``scalaron'', can be quantized, and is itself a cold dark matter candidate. Therefore, it remains to be seen whether the scalaron particles could produce the correct amount of gravitational lensing~\cite{Yadav:2018llv}.
An alternative option for ``tensor MOND" would be a second metric--like field $\tilde{g}_{\mu\nu}$:
\begin{equation}
\mathcal{S}_g \ = \ - \ \frac{M_P^2}{16\pi}\int \ d^4x\sqrt{-g}R \ - \ \frac{\tilde{M}_P^2}{16\pi}\int d^4x\sqrt{-\tilde{g}}\tilde{R} \ .
\end{equation}
If the matter Lagrangian coupled to both metrics separately, the theory would be plagued by the Boulware-Deser ghost~\cite{Boulware:1973my}~\cite{Hinterbichler:2011tt}~\cite{Hassan:2011zd}~\cite{deRham:2014zqa}. However, one can avoid the ghost by coupling matter to a certain combination of the metrics~\cite{deRham:2014naa}; it is also possible to add interaction terms between $g_{\mu\nu}$ and $\tilde{g}_{\mu\nu}$, corresponding to massive gravity~\cite{Hassan:2011zd}, and generalize this model to $f(R)$ gravity~\cite{Nojiri:2012zu}. However, unlike ordinary $f(R)$ gravity, this class of theories suffers from the same drawback as TEVES: light and gravity couple to different metrics, putting them at tension with LIGO results~\cite{Boran:2017rdn}. 
Another possibility, explored in~\cite{Platscher:2018voh}, is to couple matter only to $g_{\mu\nu}$; in this case, the long-range modifications of gravity would be produced by the interaction terms between the two metrics. However, such models cannot be called ``modified gravity" in the rigorous sense, since $\tilde{g}_{\mu\nu}$ only interacts with baryonic matter via gravity. Because of this, it would be more accurate to describe $\tilde{g}_{\mu\nu}$ as an exotic dark matter field with a non-standard coupling to gravity, and therefore, a detailed examination of this theory lies beyond the scope of this paper.
One more scenario of this type, known as ``bimetric MOND", is outlined in~\cite{Milgrom:2009gv}. Its interaction term is a function of the difference between the Christoffel symbols of the two metrics, but we shall not discuss it any further for the same reason.
\section{Fractional Kinetic Terms from Spontaneous Symmetry Breaking}\label{sec:Khoury}
A more intricate option would see the fractional power in \eqref{MG} emerge, in the non-relativistic limit, from symmetry breaking. In this fashion, it would not reflect a property of the theory itself, but of the vacuum in a non-relativistic regime. This key idea underlies Khoury's superfluid dark matter model~\cite{Berezhiani:2015bqa}, described by the scalar field Lagrangian
\begin{equation}\label{SFL}
\mathcal{L} \ = \ -\frac{1}{2}\left(|D_\mu\Phi|^2+m^2|\Phi|^2\right) \ - \ \frac{\Lambda^4}{6(\Lambda_c^2+|\Phi|^2)^6}\left(|D_\mu\Phi|^2+m^2|\Phi|^2\right)^3 \ .
\end{equation}
One can verify that in this model the energy density is properly bounded from below, since
\begin{equation}\label{a2}
\begin{gathered}
T_{00} \ = \ \frac{1}{2}\left(|D_0\Phi|^2+|D_i\Phi|^2+m^2|\Phi|^2\right)\\
\ + \ \frac{\Lambda^4}{6(\Lambda_c^2+|\Phi|^2)^6}(|D_\mu\Phi|^2+m^2|\Phi|^2)^2\left(|D_i\Phi|^2+5|D_0\Phi|^2+m^2|\Phi|^2\right)\ge 0 \ .
\end{gathered}
\end{equation}
\normalsize
In the non-relativistic limit, the scalar field can condense in gravitational wells acquiring a nonzero VEV
\begin{equation}\label{VEV}
\Phi \ = \ \rho e^{i(\theta+mt)} \ , 
\end{equation}
and then, substituting it into \eqref{SFL} and integrating out $\rho$, one obtains a fractional-power kinetic term for $\theta$ of the form $\left((\vec{\nabla}\theta)^2\right)^{3/2}$. Moreover, adding a coupling to baryons
\begin{equation}\label{BF}
\mathcal{L}_{int} \ = \ -\alpha\frac{\Lambda}{M_P}\theta\,\rho_b \ ,
\end{equation}
yields a Lagrangian similar in structure to AQUAL, from which one can derive a MOND regime. 
However, it is somewhat problematic to obtain a covariant form of \eqref{BF}, although in principle one could try to model it as a variant of ``scalar MOND", along the lines of the preceding sections
\begin{equation}
\mathcal{L}_{int} \ = \ -\alpha\frac{\Lambda}{M_P}\theta \,T_\alpha^\alpha \ ,
\end{equation}
equivalently defined with the effective metric
\begin{equation}
\tilde{g}_{\mu\nu} \ = \ e^{-2\alpha\frac{\Lambda}{M_P}\theta}g_{\mu\nu} \ .
\end{equation}

At finite temperatures, SfDM would be a mixture of superfluid and normal fluid, so that, in contrast to the standard ``scalar MOND", this setting could also account, in principle, for gravitational lensing~\cite{Hossenfelder:2018iym}. However, in the unbroken phase,
\begin{equation}
\theta \ = \ \frac{i}{2}\ln\left(\frac{\Phi^*}{\Phi}\right) \ - \ mt \ ,
\end{equation}
which is non-covariant and renders the coupling ill-defined at $\Phi=0$. The latter problem is generic in superfluid/symmetry breaking models, since Goldstone bosons usually emerge as phases of complex fields~\cite{Mistele:2019byy}. One potential way to bypass this difficulty may be to couple the superfluid not to matter density in general, but just to the baryonic matter density, in which case our Noether current would be the baryonic current, rather than the stress-energy tensor~\cite{Alexander:2018fjp}. However, such models do not qualify as ``modified gravity", so we shall not discuss them.
In addition, the phase transition mechanism is not completely clear, since \eqref{a} is equal to zero at $\Phi=0$ regardless of the presence of the gravitational potential, and is larger than zero in the broken phase, according to \eqref{VEV}, so that the broken phase ends up having a higher energy density than the unbroken one.

\section{Conclusions}\label{sec:conclusions}
In this paper I have studied possible relativistic completions of various ``modified gravity" models underlying MOND, which are all driven, one way or another, by the modified Poisson equation~\eqref{MP}. I have considered generic relativistic completions of \eqref{MG}, of the form
\begin{equation}\label{MONDL}
\mathcal{L} \ = \ \mathcal{L}_{MOND}(\phi,\ldots,\xi) \ + \ h_{\mu\nu}(\phi,\ldots,\xi)\, T^{\mu\nu} \ ,
\end{equation}
where $\mathcal{L}_{MOND}$ is the Lagrangian for the covariant ``MOND fields" $\phi$, \ldots, $\xi$, which could be arbitrary in number and could transform as scalars, vectors, or tensors. To first order in $h_{\mu\nu}$, the composite current from the MOND sector, a coupling term of the form $h_{\mu\nu}T^{\mu\nu}$ is equivalent to coupling the matter Lagrangian to an effective metric
\begin{equation}
\tilde{g}_{\mu\nu} \ = \ g_{\mu\nu} \ - \ 2h_{\mu\nu} \ ,
\end{equation}
instead of $g_{\mu\nu}$. In the special case when $\tilde{g}_{\mu\nu}=g_{\mu\nu}$ and no additional fields are present, our relativistic MOND theory is just a particular type of $f(R)$ gravity. 

One limitation of the present work is that I have restricted the analysis, for clear reasons of simplicity, to models with a single MOND field $\Psi$. When $\Psi$ is either a scalar or a vector, one can then arrive relatively simply at relativistic extensions of the MOND Lagrangian where the fractional power of the kinetic term remains manifest. I have deemed it reasonable to impose some constraints on these models. The first, key constraint, is that the energy of field configurations be bounded from below, while a second, equally reasonable constraint, is the absence of superluminal propagation modes. The comparison with data introduces however further constraints, which are related to gravitational lensing effects. I have shown that a general scalar field theory of the form $\chi^\lambda$, where $\chi$ is a quadratic combination of field derivatives, is bounded from below only if $\lambda$ is positive and not an even number, while for a vector field, $\lambda$ should be greater than $\frac{1}{2}$, and again not an even number. Conversely, for a kinetic term of the form $|\chi|^{\lambda-1}\chi$, which allows to define the theory for negative values of $\chi$, the condition would be $\lambda\ge\frac{1}{2}$ for both scalars and vectors, and even values of $\lambda$ would be allowed. In addition, I have shown that, for a vector field $u$, $\chi$ should have the Maxwell form $\nabla_\mu u_\nu-\nabla_\nu u_\mu$, which is not the case for the covariant emergent gravity theory proposed in~\cite{Hossenfelder:2017eoh} as a relativistic extension of the entropic gravity of~\cite{Verlinde:2016toy}. However, the Bianchi identity implied by a Maxwell-type kinetic term makes a MOND-type coupling to matter impossible. As a result, a MOND--like behavior would entail a complicated and rather baroque coupling. While superluminal modes are absent in Maxwell-type vector theories for $\lambda\ge1$, they are present in scalar theories for $\lambda>1$, although it was argued in~\cite{Bruneton:2006gf} that they do not violate causality, because the equations of motion admit nonetheless a well-posed Cauchy problem. Nonetheless, ``scalar MOND" cannot reproduce gravitational lensing, since the scalar field is just a conformal factor of the metric and therefore does not affect the Maxwell field.

The tensor case is different, since the tensor kinetic term is a function of the Ricci scalar, and therefore one cannot recover a fractional kinetic term in the non-relativistic limit. However, one can attempt to reproduce the MOND potential for a point mass. I have demonstrated that such a solution actually exists in $f(R)=R^\lambda$ theories with $\lambda>1$, but it is problematic to connect it, at large distances, to a Schwarzschild solution at smaller ones. This is largely due to the fact that the family of solutions is too broad, and there is no reason for the system to ``prefer" one of them over another. In addition, since $f(R)$ gravity is equivalent to a scalar-tensor theory that couples to matter via a conformal factor (as is also the case with ''scalar MOND"), a generic $f(R)$ theory cannot account for additional effects of gravitational lensing. I have also stressed, however, that it might be possible to reproduce the cold dark matter phenomenology, including the lensing observations, with a quantized scalar field. Another option, a bimetric theory in which matter and gravity couple to different metrics, does not fit well with the LIGO gravitational wave observations, according to which light and gravity should travel on the same geodesics~\cite{Boran:2017rdn}.
Based on these simple examples, I have gathered some evidence to the effect that modified gravity theories must have the form \eqref{MONDL}, but then they either do not reproduce gravitational lensing (as is the case when $h_{\mu\nu}\propto g_{\mu\nu}$, for example, in scalar MOND or in $f(R)$ gravity) or produce different geodesics for gravity and matter, contradicting the LIGO phenomenology~\cite{Boran:2017rdn}. This considerably limits the potential lessons of purely ``modified gravity" theories in connection with MOND phenomenology. Under certain assumptions (such as Lorentz invariance and absence of modified inertia), the ``conformal rescaling" appears the only viable way of modifying the gravitational coupling to matter. In this context, gravitational lensing observations could only be explained by \emph{actual} gravitational interactions of some new particles/fields with light.

All the preceding arguments do not rule out hybrid models, which combine a cold dark matter-like component with ``conformal rescaling"-like modifications of gravity, treating them either as separate entities or as two different manifestations of the same phenomenon. Two notable examples of the latter setup are the aforementioned $f(R)$ gravity and the superfluid dark matter model, which rests on a scalar field Lagrangian that can undergo spontaneous symmetry breaking~\cite{Berezhiani:2015bqa}. In the unbroken phase, the field behaves like cold dark matter, while in the broken phase it produces a phonon--driven interaction that mimics MOND. In this context, the fractional kinetic term only emerges around the non-relativistic vacuum of the broken phase. I have shown that the energy density is bounded from below in SfDM, and that, while it was already known~\cite{Hossenfelder:2018iym} to yield gravitational lensing, it is problematic to introduce a covariant coupling to baryons within this theory. Moreover, the phase transition mechanism is unclear, because the energy density of the system in the broken phase is \emph{higher} than in the unbroken one. One could summarize these findings saying that the hybrid models discussed in this paper (SfDM and $f(R)$) are incomplete in their present form, and it remains to be seen whether a satisfactory completion is possible. Likewise, the results of our analysis do not exclude theories that mimic aspects of MOND without having the Lagrangian structure of eq.~\eqref{MONDL}. Notable examples of such ``fake modified gravity" theories include exotic dark matter fields that have non-standard couplings to gravity~\cite{Platscher:2018voh}~\cite{Milgrom:2009gv} and fields that couple to the baryonic charge, rather than to the matter density~\cite{Alexander:2018fjp}.

Since two of the aforementioned results (namely, the LIGO constraints and the $f(R)$ interpolation problem between small and large radii) are generic and do not depend on the specific form of the potential, they have far going implications for long-range/infrared modifications of gravity in general, and limit considerably the viable options. The lessons that we have gathered from energy bounds for scalars and vectors, in our opinion, are useful inputs to guide the search for more general effective field theories.
In~\cite{sz} we shall explore an alternative option that can in principle account, at least in part, for the flattening of rotation curves. It rests solely on classical physics: mass distributions for dark matter that bulge away from galactic planes to an extent $\Delta$ are natural sources of quasi-logarithmic potentials within distances of order $\Delta$. 

\acknowledgments %

I am grateful to A. Sagnotti for support and suggestions, and for his feedback on the manuscript, and to M.~Cadoni, K.~Mkrtchyan and M.~Tuveri for useful comments. It is a pleasure to acknowledge the kind hospitality of the Deutsches Elektronen-Synchrotron (DESY) center in Hamburg, where most of this work was carried out. The author was supported in part by Scuola Normale Superiore, by INFN (IS CSN4-GSS-PI) and by the MIUR-PRIN contract 2017CC72MK\_003.


\bibliographystyle{JCAP}

\end{document}